\documentstyle [a4,12pt]{article}

\begin{document}
\begin{center}
{\large \bf ABOUT LIMIT MASSES OF ELEMENTARY PARTICLES}\\
 \vspace{0,5cm}
 {\bf Umida R.Ibadova}\\
 \vspace{0,5cm}
 Department of Theoretical Physics and
Computer Sciences, of Samarkand State University, Uzbekistan. 15,
University blvd, 703004 Samarkand, Uzbekistan.\\
 ibrumida@yahoo.com
\end{center}
\vspace{0.8cm}

{\bf Abstract}\\
The simple examples of spontaneous breaking of various symmetries
for the scalar theory with fundamental mass have been
considered\footnote{This research was supported by the Belgian
National Fund for Scientific Research}. Higgs' generalizations on
"fundamental mass" that was introduced into the theory on a basis of
the five-dimensional de Sitter space. The connection among
"fundamental mass", "Planck's mass" and "maximons" has been found.
Consequently, the relationship among G- gravitational constant and
other universal parameters can
be established.\\

{\bf Key words:}  Quantum field theory, fundamental mass,spontaneous
breaking, ultrahigh
energies.\\

{\bf 1. Introduction}\\

The concept of mass having its root from deep antiquity (including
Galileo's Pisans experiment, theoretical research of the connection
of a mass with the Einstein's energy etc.) still remains
fundamental. Every theoretical and experimental research in
classical physics and quantum physics associated with mass is a step
to the discernment of Nature. Besides mass, the other fundamental
constants such as Plank's constant and the speed of light also play
the most important role in the modern theories. The first one is
related to quantum mechanics and the second one is related to the
theory of relativity. Nowadays the properties and interactions of
elementary particles can be described more or less adequately in
terms of local fields that are affiliated with the lowest
representations of corresponding compact groups of symmetry. It is
known that mass of any body is composed of masses of its comprising
elementary particles. The mass of elementary particles is the
Kasimir's operator of the noncompact Paunkare group, and those
representations of the given group, that are being used in Quantum
Field Theory (QFT), and it can take any values in the interval of  .
Two particles, today referred to as elementary particles, can have
masses distinct one from another by many orders. For example, the
vectorial bosons with the mass of $\sim10^{15}GeV$ take place in GUT
modules, whereas the mass of an electron is only $\sim 10^{-4} GeV$.
Formally, the standard QFT remains logical in a case when the mass
of particles can be compared with the mass of automobiles. The
modern QFT does not forbid such a physically meaningless
extrapolation. Perhaps, it is a principal defect of the theory?  In
1965 Markov put forward a hypothesis \cite{1}, according to which
the spectrum of masses of elementary particles must jump into
discontinuity on the "Plank's mass", well-known universal constants,
and G-gravitational constant take place in this expression. Markov
named particles of a limiting mass as $"maximums"$. In works
V.G.Kadyshevsky and his students\cite{2,3} in the quantum theory of
a field on a strict mathematical basis the parameter " fundamental
mass " which alongside with parameters of the standard quantum
theory to play an essential role in physics at high energy is
entered. In the given work, we have strived for the connection of
the universal constants $(\hbar, c, G)$, "Plank's mass",  "maximon"
and "fundamental mass" on a basis of the spontaneous breaking of
symmetry.\\

{\bf 2. On Quantum Field Theory with a new parameter fundamental
mass}\\

Here we will cite briefly the major ideas of quantum field theory
(QFT), in which the four-dimensional momentum space is a constant
curvature space with a large radius M.

The construction of the sequential Quantum Field Theory (QFT) with a
new, universal scale in the superhigh energy region- Fundamental
Mass M emanates to V.G. Kadyshevsky's work \cite{2} and his students
\cite{3}. The parameter M is called fundamental mass and its inverse
fundamental length. The fundamental mass fixes a new universal scale
of theory in the high-energy region. The standard QFT is recovered
in the "flat limit" . Recall that fundamental mass M is a new
hypothetical parameter of mass dimension, which should be as
universal as   or Newtonian gravitational constant $G$ and serve as
characteristic scale in the region of high energies. A key role in
the approach developed belongs to the five-dimensional configuration
representation. Being four-dimensional in its essence the theory
admits a specific local Lagrangian formulation in which the
dependence of fields on auxiliary fifth coordinate is local too.
Internal symmetries in this formalism generate gauge transformations
localized in the same five-dimensional configuration space.  The de
Sitter space has a constant curvature. Depending on its sign there
are two possibilities The de Sitter space has a constant curvature.
Depending on its sign there are two possibilities

\begin{equation} \begin{array}{ccc}
p^{2}_{0}- p^{2}_{1}- p^{2}_{2}- p^{2}_{3}+ p^{2}_{5}\equiv g^{KL}P_{K}P_{L}=M^{2};\\
\hspace{1cm}\small {K,L=0,1,2,3,5}\\
(positive\  curvature: \quad
g^{00}=-g^{11}=-g^{22}=-g^{33}=g^{55}=1) \end{array}
\end{equation}
\begin{equation}
\begin{array}{ccc}
p^{2}_{0}- p^{2}_{1}- p^{2}_{2}- p^{2}_{3}- p^{2}_{5} \equiv g^{KL}P_{K}P_{L}=-M^{2};\\
\hspace{2cm}\small {K,L=0,1,2,3,5}\\
(negative\  curvature: \quad
g^{00}=-g^{11}=-g^{22}=-g^{33}=-g^{55}=1).
\end{array}
\end{equation}

It is natural that QFT based on momentum representation of the form
(1)-(2) must predict new physical phenomena at energies  . In
principle, the parameter M may turn out to be close the Planck mass
GeV. Then, the new scheme should include quantum gravity. The
standard QFT corresponds to the "small" 4-momentum approximation and
which formally can be performed by letting   ("flat limit").  The
formulation of QFT with fundamental mass discussed in this paper is
based on the quantum version of the de Sitter equation (2), i.e. on
the five-dimensional field equation
\begin{equation}
\left[\frac{\partial^2}{\partial x^{\mu}\partial x_{\mu}}-
\frac{\partial^2}{\partial x^2_5}-\frac{M^2c^2}{\hbar^2}\right]
\Phi(x,x^5)=0
\end{equation}
$$
\small{\qquad\qquad\qquad \mu =0,1,2,3.}
$$

All the fields independently of their tensor dimension must obey (3)
since similar universality was inherent in the "classical" prototype
of (3)- de Sitter p-space (2). As applied to scalar, spinor, vector
and other fields we shall write down the five-dimensional wave
function $\Phi(x,x^5) $ in the form $\varphi (x,x^{5} ), $ $\psi
_{\alpha } (x,x^{5} )$ and $A_{\mu } =(x,x^{5} )$   and  the
decision (3) in view of the relevant Cauchy problem, which is
correctness on $x^{5}$, looks like:

\begin{equation}\Phi(x,x^5)\leftrightarrow \left(
\begin{array}{c}
\Phi(x,0) \\
\partial \Phi(x,0)/\partial x^{5} .
\end{array}
\right)\equiv\left(\begin{array}{c}
\Phi(x) \\
\chi(x)
\end{array}
\right).
\end{equation}

 In other words, the statement that to
each field in the 5-space there corresponds its wave function
$\Phi(x,x^5)$ obeying (3), implies that each of these fields in the
usual space-time is described by the wave function with a doubled
number of components. Then, it is natural to assume that the initial
date (4) obey the Lagrangian equations of motion following from the
action principle
\begin{equation}
S=\int d^{4}x L\left[ \Phi (x,0),{\partial \Phi (x,0)\over \partial
x^{5}}\right].
\end{equation}
A key role in the approach developed belongs to the five-dimensional
configuration representation. Being four-dimensional in its essence
the theory admits a specific local lagrangian formulation in which
the dependence of fields on an auxiliary fifth coordinate is local
too. Internal symmetry's in this formalism generate gauge
transformations localized in the same five-dimensional configuration
space.\\

{\bf 3. Calculations}\\

Let's consider simple examples of spontaneous breaking of various
symmetries for the scalar theory with a fundamental mass. The real
scalar field, whose Lagrangian can be obtained from the beneath
expression:
  \begin{equation}
  L(x,M)=\frac{1}{2} [[ \frac{\partial \Phi (x)}{\partial
  x_{n}}]^{2}+m^{2}\Phi^{2}(x)+M^{2}[\chi (x)-\cos\mu\Phi(x)]^{2}]
\end{equation}
Using (4) and it can be expressed as follows:
 \begin{equation}
  L(x,M)=\frac{1}{2} [ \frac{\partial \varphi(x)}{\partial
  x_{\mu}}]^{2}- \frac{1}{2}m^{2}\varphi(x)^{2}-\frac{1}{2}M^{2}(\chi
  (x)-\cos\mu\varphi(x))^{2}-\chi(x)U(\varphi(x)
\end{equation}

Here $\cos\mu=\surd\overline{1-\frac{m^{2}}{M^{2}}}$  where $m$ -
mass of particles, described by fields $\varphi, U(\varphi )$ -an
unknown function characterizing interactions among these particles.
Is it possible to select the interaction $L_{int}$ between the
fields $\varphi$ and $\chi$ in order that under the exclusion of the
field $\chi$, the Higgs potential is appropriate to the field
$\varphi$ ?

 The free Lagrangian is invariant relatively the
transformations $\varphi\rightarrow -\varphi$ and $\chi\rightarrow
-\chi$. It is necessary to require that $U(\phi)\rightarrow
U(-\phi)$ , i.e. $U(\phi)$-an odd function of $\varphi$, an action
for (7) will be written as:
  \begin{equation}
  S(x,M)=\int{\{\frac{1}{2} [ \frac{\partial \varphi(x)}{\partial
  x_{\mu}}]^{2}- \frac{1}{2}m^{2}\varphi(x)^{2}-\frac{1}{2}M^{2}(\chi
  (x)-\cos\mu\varphi(x))^{2}+\chi(x)U(\varphi(x)\}}d\varphi
\end{equation}
Taking a derivative of (8) on $\chi$  we obtain:
\begin{equation}
\chi=\cos\mu\varphi+\frac{U(\varphi)}{M^{2}}
\end{equation}
Substituting (9) to (7), we have:
\begin{equation}
  L_{tot}=\frac{1}{2} [ (\frac{\partial \varphi(x)}{\partial
  x_{\mu}})^{2}- m^{2}\varphi(x)^{2}+\frac{U^{2}(\varphi)}{M^{2}}
  +2U(\varphi(x))\cos\mu\varphi(x))]
\end{equation}
that is invariant relatively $\varphi\rightarrow -\varphi $. We know
from spontaneous breaking of the discrete symmetry for the usual
scalar field, that the Higgs potential will have a view:
\begin{equation}
V(\varphi)=-\frac{1}{2}m^{2}\varphi(x)^{2}+\frac{1}{4}{\lambda^{2}\varphi(x)^{4}}
\end{equation}
where  $\lambda$ -a mass less constant, characterizing interactions
among particles.

Let's find a form of the function $U(\varphi)$ , in order for the
Higgs potential to figurate in (10). Let's consider the Lagrangian
(10) under $m\rightarrow -m$, then
$$
\cos\mu=\sqrt{1-\frac{m^{2}}{M^{2}}}\rightarrow\sqrt{1+\frac{m^{2}}{M^{2}}}
$$
The potential energy (11)
will have a view:
\begin{equation}
V(\varphi)=-\frac{1}{2}m^{2}\varphi(x)^{2}-\frac{U^{2}}{2
M}-U(\varphi)ch\mu\varphi
\end{equation}
Comparing (12) and (11) for $U(\varphi)$  we have two different
roots (real and imaginary) under $$\varphi^{2}<\frac{2
M^{2}ch^{2}\mu}{\lambda^{2}}$$ and $$U=-M^{2}ch^{2}\mu\varphi$$
under $$\varphi^{2}=\frac{2 M^{2}ch^{2}\mu}{\lambda^{2}}$$ This
results in this that $
L_{tot}(\varphi)_{Higgs}=L^{0}_{maximon}(\varphi) $ i.e.
\begin{equation}
L^{0}_{maximon}(\varphi)=\frac{1}{2}(\frac{\partial
\varphi}{\partial x_{\mu}})^{2}-\frac{1}{2}M^{2}\varphi^{2}
\end{equation}
Now let's consider a case when $$ L_{tot}(\varphi)=-
\frac{\lambda^{2}}{4} \varphi^{2}\chi^{2}$$
Under $m\rightarrow -m$,
the Lagrangian (7) will have a view:
\begin{equation}
L_{tot}(\varphi)=\frac{1}{2}[(\frac{\partial \varphi}{\partial
x_{\mu}})^{2}-M^{2}sh^{2}\mu^{`}\varphi^{2}-M^{2}(\chi-ch\mu^{`}\varphi)^{2}-
\frac{\lambda^{2}}{2}\varphi^{2}\chi^{2}]
\end{equation}

Taking a derivative of (14) $\chi$ on  $$ \chi=
\frac{M^{2}ch\mu^{`}\varphi}{M^{2}+\frac{\lambda^{2}}{2}\varphi^{2}\chi^{2}}$$
we obtain:

\begin{equation}
L_{tot}(\varphi)=\frac{1}{2}[(\frac{\partial \varphi}{\partial
x_{\mu}})^{2}+M^{2}sh^{2}\mu^{`}\varphi^{2}-\frac{\lambda^{2}}{2}\varphi^{2}(
\frac{\lambda^{2}\varphi^{2}}{2M^{2}}+1)\frac{M^{4}ch\mu^{`}\varphi^{2}}{(M^{2}+
\frac{\lambda^{2}\varphi^{2}}{2})^{2}}]
\end{equation}

This is one of the Higgs' generalizations on a fundamental mass. We
will obtain the usual Higgs' Lagrangian from (15) under
$M\rightarrow\infty$:

\begin{equation}
\lim_{M\rightarrow\infty}
L_{tot}(\varphi)=\frac{1}{2}[(\frac{\partial \varphi}{\partial
x_{\mu}})^{2}+m^{2}\varphi^{2}-\frac{\lambda^{2}\varphi^{4}}{2}]
\end{equation}

If taking into account in (11) the strong connection
$\lambda\rightarrow\infty$ then we come to a Lagrangian:

$$
L_{tot}(\varphi)=\frac{1}{2}(\frac{\partial \varphi}{\partial
x_{\mu}})^{2}- M^{2}\varphi^{2},
$$ i.e.
$$
L_{tot}(\varphi)_{Higgs}=L^{0}_{maximon}(\varphi)
$$
Spontaneous breaking of the global $U(1)$-symmetry results in:
$$
L_{tot}=|\frac{\partial \varphi}{\partial
x}|^{2}-m^{2}|\varphi|^{2}-M^{2}|\chi-\cos\mu\varphi|^{2}
-\frac{\lambda^{2}}{2}|\chi|^{2}|\varphi|^{2}=
$$
\begin{equation}
=|\frac{\partial \varphi}{\partial
x}|^{2}-M^{2}|\varphi|^{2}-M^{2}|\chi|^{2}+M^{2}\cos\mu|\chi\overline{\varphi}+
\varphi\overline{\chi}|-\frac{\lambda^{2}}{2}|\chi|^{2}|\varphi|^{2}
\end{equation}
 under $m\rightarrow i m$, then
\begin{equation}
L_{tot}=|\frac{\partial \varphi}{\partial
x}|^{2}-M^{2}|\varphi|^{2}-M^{2}|\chi|^{2}+M^{2}ch\mu^{`}|\chi\overline{\varphi}+
\varphi\overline{\chi}|-\frac{\lambda^{2}}{2}|\chi|^{2}|\varphi|^{2}
\end{equation}

This Lagrangian differs from (17) by a sign before $m^{2}$ , but it
is still invariant relatively the group of global transformations:
\begin{equation} \begin{array}{ccc}
\varphi(x)\rightarrow\varphi(x)=\exp(igs)\varphi(x), \qquad \qquad
\varphi(x)^{*}\rightarrow\varphi(x)^{*}=\exp(-igs)\varphi(x)^{*}\\

\chi(x)\rightarrow\chi(x)=\exp(igs)\chi(x), \qquad \qquad
\chi(x)^{*}\rightarrow\chi(x)^{*}=\exp(-igs)\chi(x)^{*}
\end{array}
\end{equation}

Taking the derivative of $L_{tot}(x)$ on $\chi$  and
$\overline{\chi}$ , we will find the equation of motion for
$\overline{\chi}$  and $L_{tot}(x)$  respectively
$$-M^{2}\overline{\chi}+M^{2}ch\mu^{`}\overline{\varphi}-
\frac{\lambda^{2}}{2}|\varphi|^{2}\overline{\chi}=0 $$ and
$$-M^{2}\overline{\chi}+M^{2}ch\mu^{`}\overline{\varphi}-
\frac{\lambda^{2}}{2}|\varphi|^{2}\chi=0 $$
From here we get:
$$
\overline{\chi}=\frac{ch\mu^{`}\overline{\varphi}}{1+\frac{\lambda^{2}}{2M^{2}|\varphi|^{2}}}
$$ and
$$
\chi=\frac{ch\mu^{`}\overline{\varphi}}{1+\frac{\lambda^{2}}{2M^{2}|\varphi|^{2}}}
$$

\begin{equation}
L_{tot}=|\frac{\partial \varphi}{\partial x}|^{2}-V(\varphi)
\end{equation}
where $V(\varphi)$-Higgs potential
$$ V(\varphi)=M^{2}|\varphi|^{2}-\frac{M^{2}ch^{2}\mu^{`}|\varphi|^{2}}{1+
\frac{\lambda^{2}}{2M^{2}}|\varphi|^{2}}
$$
 under $|\varphi|^{2}=\frac{2M^{2}}{\lambda^{2}}(ch\mu^{`}-1)$ has its
$V_{min}(\varphi)=-\frac{2M^{4}}{\lambda^{2}}(ch\mu^{`}-1)^{2}$. In
the flat limit $M\rightarrow\infty$, (20) will have a usual view. If
we write $V(\varphi)=V_{min}(\varphi)$, then

\begin{equation}
V_{New}(|\varphi|)=\frac{\lambda^{2}}{2}\frac{[|\varphi|^{2}-\frac{h^{2}}{2}]^{2}}{1+
\frac{\lambda^{2}}{2M^{2}}|\varphi|^{2}}
\end{equation}
where $\frac{h^{2}}{2}=\frac{2M^{2}}{\lambda^{2}}(ch\mu^{`}-1)$
under $M\rightarrow\infty$ equals $\frac{m^{2}}{\lambda^{2}}$.

Finally, we obtain
\begin{equation}
L_{tot}(\varphi)=|\frac{\partial\varphi}{\partial x}|^{2}-
\frac{\lambda^{2}}{2}\frac{[|\varphi|^{2}-\frac{h^{2}}{2}]^{2}}{1+
\frac{\lambda^{2}}{2M^{2}}|\varphi|^{2}}
\end{equation}

 The system, described by the Lagrangian (18), has the
spontaneous broken $U(1)$-symmetry. Now a point is not appropriate
to the minimum of energy $\varphi(x)=\overline{\varphi}^{*}(x)$ ,
but any point on a circle of the radius
$$
R=\sqrt{2}\frac{M^{2}}{\lambda}\sqrt{ch\mu^{`}-1}
$$
 Let's write $\varphi(x)$ in a view of the real and imaginary parts. As a stable
vacuum, we can choose any state, lying on a circle of the radius of
R, i.e. all the states are equivalent due to invariation relatively
transformations (19).

 Let's pick out the magnitude of a gauge phase
to be $\beta=0$, identical for the whole world, and let's write
$\varphi$ in a view
\begin{equation}
\varphi(x)=\frac{1}{\sqrt{2}} (h+\varphi_{1}(x)+i\varphi_{2}(x))
\end{equation}
Substituting (23) into (22), we obtain
$$
L_{tot}(\varphi)\Rightarrow L_{tot}(\varphi_{1},\varphi_{2})=
$$
\begin{equation}
=\frac{1}{2}[\frac{\partial\varphi_{1}(x)}{\partial x_{\mu}}]^{2}+
\frac{1}{2}[\frac{\partial\varphi_{2}(x)}{\partial x_{\mu}}]^{2}-
\frac{\lambda^{2}}{8}\frac{[\varphi_{1}^{2}+2h\varphi_{2}^{2}]^{2}}{1+\frac{\lambda^{2}}{16
M^{2}}[(h +\varphi_{1})^{2}+\varphi_{2}]^{2}}
\end{equation}

The Goldstone scalar mass less particle $\varphi_{2}$ has appeared
as a result of the spontaneous breaking of the symmetry, and a real
scalar particle with a mass
\begin{equation}
m_{1}=\frac{\lambda h}{[1+\frac{\lambda^{2}h^{2}}{16 M^{2}}]^{1/2}}
\end{equation}
Under $M\rightarrow\infty$ from (25) we have $m_{1}=\sqrt{2m}$. If
we consider the real scalar particle as a maximon, then from (25) we
obtain:
$$
m_{1 maximon}=\frac{2M}{[\sqrt{2}+125]^{1/2}}
$$

{\bf Acknowledgements}\\

Author would like to thank sincerely grateful to V. Gogokhia, M.
Ausloos, M.Musakhanov and B.Yuldashev for useful arguing of
outcomes, constant attention and help with a spelling of the given
work. This work was supported by the Belgium Grant Scientific
research fellowship.

\end{document}